\documentstyle[12pt]{article} 

\newcommand{\beq}{\begin{equation}}
\newcommand{\eeq}{\end{equation}}
\newcommand{\bea}{\begin{eqnarray}}
\newcommand{\eea}{\end{eqnarray}}
\newcommand{\bean}{\begin{eqnarray*}}
\newcommand{\eean}{\end{eqnarray*}}

\newcommand{\gapprox}{\lower.7ex\hbox{$\;\stackrel{\textstyle>}{\sim}
\;$}}
\newcommand{\lapprox}{\lower .7ex\hbox{$\;\stackrel{\textstyle
<}{\sim}\;$}}
\def\Br{Br}%
\def\thefiglist#1{\section*{Figure Captions\markboth
{FIGURE CAPTIONS} {FIGURE CAPTIONS}}\list
{Figure \arabic{enumi}}
{\settowidth\labelwidth{Figure #1.}\leftmargin\labelwidth
\advance\leftmargin\labelsep
\usecounter{enumi}.}
\def\newblock{\hskip .11em plus .33em minus -.07em}
\sloppy}

\begin{document}
\begin{titlepage}

\begin{flushright}
\begin{tabular}{l}
hep-ph/9708265\\
RAL-97-036 \\
FERMILAB-PUB-97/232-T\\
HUTP97/A-040
\end{tabular}
\end{flushright}

\vskip 0.3cm
\begin{center}
\Large
{\bf Gluonic Hadrons and Charmless $B$ Decays}

\vskip 0.75cm

\large
Frank E. Close \\
{\small Rutherford Appleton Laboratory,
Chilton, Didcot, Oxon OX11 OQX, UK}
\vskip 0.30cm

Isard Dunietz \\
{\small Fermi National Accelerator Laboratory,
P.O. Box 500, Batavia, IL 60510, USA}
\vskip 0.30cm

Philip R. Page \\
{\small Thomas Jefferson National Accelerator Facility,
12000 Jefferson Avenue, \\ Newport News, VA 23606, USA}
\vskip 0.30cm

Sini\v{s}a Veseli \\
{\small Fermi National Accelerator Laboratory,
P.O. Box 500, Batavia, IL 60510, USA}
\vskip 0.30cm

Hitoshi Yamamoto \\
{\small Harvard University, 42 Oxford St., Cambridge, MA 02138, USA }

\end{center}
\vskip 0.5cm

\begin{abstract}
Hybrid charmonium with mass $\sim$4 GeV could be produced via a
$c\overline{c}$ color-octet component in $b\rightarrow  c\overline{c} s$.
These states could be narrow and could have a significant branching ratio to
light hadrons, perhaps enhanced by glueballs. Decays to gluonic hadrons could
make a sizable contribution to $B\rightarrow$ no charm decays. Experimental
signatures and search strategies are discussed.

\end{abstract}

\end{titlepage}

\section{Introduction}
Intensive studies of $B$-meson decays are now underway and will soon be
improved further with the emergence of $B$-factories that will lead
to orders of magnitude increase in statistics.
While the primary emphasis of these developments is in studying CP-violation
and seeking evidence of physics beyond the standard model, we propose
here that they may provide a powerful tool to search for
a missing piece of the standard model, namely the predicted existence of hybrid
(quark--antiquark--gluon) mesons and glueballs. We point out that there could
be a sizable production of  $c\bar cg$ hybrids (or hybrid charmonia, denoted
hereafter as $\psi_g$) and other gluonic hadrons in $B$ decays which may be
experimentally observable.

Our motivation is based on the following, {\it a priori} independent, features
of data and theory:
\begin{itemize}
\item[(i)]
  CLEO has recently reported large values of $\Br(B \to K\eta')$ and
  $\Br(B \to \eta' X, P_{\eta'} > 2$ GeV)~\cite{smith, CLEOetap}.
\item[(ii)]
  The branching ratio $\Br(b\rightarrow$ no open charm)
  appears to be about a factor of 3 larger than expected
  ~\cite{dun97}.
\item[(iii)]
  The CKM--favored decay $b\rightarrow
  c\overline{c}s$ is predicted to
  produce the $c\overline{c}$ pair in a
  colour octet configuration~\cite{dun97}.
\item[(iv)]
  Decays $\psi_g\to D^{(*)} D^{(*)}$\footnote{Hereafter,
  the notation $D^{(*,**)} D^{(*,**)}$ implies
  $D^{(*,**)} \overline D^{(*,**)}$ or
  $\overline D^{(*,**)} D^{(*,**)}$.}
  may be suppressed by a selection rule~\cite{page97sel,cp95}.
\end{itemize}
As will be discussed later, item (iii) might enhance the
formation of $\psi_g$ states,
some of which are predicted to
be in the $4.2\pm 0.2$ GeV region~\cite{hy1}-\cite{ono}.
Furthermore, item (iv) indicates that
if $\psi_g$ states occur below the $D D^{**}$
threshold ($\sim  4.3$ GeV), there is a possibility
that they will cascade into conventional $c\overline{c}$ states as
$\psi_g (c\bar cg)\rightarrow
(c\overline{c}) (gg) \rightarrow (c\overline{c})$ + light hadrons~\cite{fc95},
or that they will directly decay via $\psi_g\rightarrow
n g\rightarrow$ light hadrons (where $n \geq 2$)~\cite{dun97,hawaii} including
resonant glueball
enhancement. The latter process
might explain the enhanced $\Br(b\rightarrow$ no open charm)
[item (ii)], and the gluonic content in the final state may contribute to the
$\eta'$ enhancement, for  example through
rescattering of $K^{(*)} \psi_g$ intermediate states: $B \to K^{(*)} \psi_g$
$\to K \eta',\; \eta' X$ [item (i)].  Not only $\psi_g$ but also other gluonic
hadrons  may be produced in $B$ decays at rates enhanced by non--perturbative
effects over traditional
expectations~\cite{simmawyler}.

In the following, we will briefly summarize the spectroscopy of
gluonic hadrons, discuss their production and decay in the $B$
meson environment, and propose
experimental search strategies.

\section{Hybrid charmonium spectroscopy}

A rich spectroscopy  of hybrid charmonium is predicted by lattice
gauge theory, flux--tube,
bag models and QCD sum
rules.    These include exotic $J^{PC} = 0^{\pm -}, 1^{-+}, 2^{+-}$ as well as
conventional $J^{PC} = 0^{-+}, 1^{--},$ etc.
Lattice gauge theory with heavy quarks predicts $ 4.04 \pm 0.03$
GeV for the spin-averaged masses of
$J^{PC} = 1^{--}, 1^{++}, (0, 1, 2)^{-+}$, $(0, 1, 2)^{+-}$
in the quenched approximation~\cite{hy1}.\footnote{Unquenching is estimated to
raise the
mass by $0.15$ GeV.}  Flux--tube models inspired by the
lattice predict 4.2 -- 4.5 GeV \cite{hy2}  within an adiabatic separation of
quark and
flux--tube motion.  More recent numerical solutions of the model find
4.1 -- 4.2 GeV \cite{bcs}.  An adiabatic bag model calculation
found $\simeq$ 4 GeV with
an overall uncertainty of $\simeq$ 200 MeV \cite{hy3,ono}.  QCD sum rules are
less clear, solutions spanning 4.1 -- 5.3 GeV \cite{sum}.

There is some support for these $\sim 4$ GeV mass
predictions when one compares with the light
quark sector where the flux--tube models find 1.8 -- 1.9 GeV
\cite{hy2,bcs} in accord
with the $\pi_g$(1800) candidate \cite{cp95,cpcp}.  Lattice QCD provides some
indication that the $J^{PC} = 1^{-+}$ may be the lightest of the exotic
states. A candidate for a $J^{PC} = 1^{-+}$ hybrid has
been reported~\cite{bnl95} in the predicted region of $2.0\pm 0.2$
GeV~\cite{hy1}. Thus the emerging hints from the light quark
sector and the stability of predictions within QCD inspired models and lattice
QCD suggest that $\psi_g$ excitations could arise in a kinematically accessible
region in the $b\to c\bar c s$ decay.

\section{Hybrid production}

A variety of hybrid excitations can be produced in $B$ decays.\footnote{The
production
of hybrid $D_g$ states $[\equiv c \overline q g]$ could play a non--negligible
role in non--leptonic and
semi--leptonic $B-$decays; this contrasts with $D$ or $D_s$ decays where
$K_g$ or $\pi_g$ may mix with the charmed mesons \cite{cl96}. A moderate
production of
$D_g$ in semi--leptonic $B$ decays $\overline B\rightarrow c\overline{q}g+l\nu$
could solve the puzzle of why
exclusive $\overline B \to (D, D^*, D^{**}) l \nu$ transitions do not saturate
the inclusive semi--leptonic branching ratio \cite{semileptonic,alephsl}.}
The production of $\psi_g$ may be significant since in the
$b\to c\bar cs$ transition the $c\bar c$ pair
is dominantly produced in colour octet~\cite{dun97,fc95,hawaii}
which may strongly couple to the $c\bar c$ pair in $\psi_g$.

The direct $\psi$ production in $b$ decays is
(0.82 $\pm$ 0.08)\%~\cite{browder},
and is not well understood theoretically.
It appears to be enhanced somewhat over estimates based on the
assumption~\cite{buras} of color--suppressed
factorization~\cite{desh,braaten}.
The factorization assumption allows for the direct decay
$b\to s\,\chi_{c1}$ but not for
$b\to s\,\{\chi_{c0}, \chi_{c2}, h_c\}$.
Thus, decisive observations
of such modes would either necessitate a non--factorizable (such as a direct
color--octet)
contribution~\cite{braaten} or a feed-down from higher mass metastable
states~\cite{fc95}, which are expected to cascade
to other charmonia observed in $B$ decays as well.

If we take the production of $\chi_{c2}$ as a measure of the colour
octet production in $B$ decays, we expect from the CLEO datum
$\Br(B\rightarrow \chi_{c2}X)=0.0025\pm 0.0010$~\cite{browder}
that $\psi_g$ should be produced
competitively at $\Br\geq 0.1\%$. The sum total of  $\psi_g$ for all $J^{PC}$
could be ${\cal O}(1\%)$, a significant contribution to the ``non-charm" $B$
decays though not saturating them. If their production is to saturate these
events, then their combined $\Br$ should be of
${\cal O}(10\%)$ and their preferred
decays ought to be to light hadrons.
If $\Br(B \to \psi_g ({\rm all}\;J^{PC}) X)\sim {\cal O}(1\%)$, then
$\Br(\psi_g\to (c\bar c) X) = {\cal O}(10-100\%)$ is
still consistent with the measured $\Br(B \to (c \overline c) + X)$. If
$\psi_g$ are produced at ${\cal O}(10 \%)$, saturating the missing ``non-charm"
decays, then cascades to
$(c \overline c)$ must be a small fraction of the total.
Unless some special mechanism causes $\psi_g$ to cascade into $h_c$ or other
undetected conventional $(c\bar c)$ states, the measurements on inclusive
$\eta_c, \psi$ or $\chi_c$
production constrain the  product of branching ratios
$\Br(B\rightarrow \psi_gX) \times \Br(\psi_g\rightarrow (c\overline{c}) X)$.

We note that the $c\bar c$
invariant mass distribution for the quark-level
$V-A$ transition $b\rightarrow c\overline{c} s$
peaks in the $3-3.7$ GeV mass range~\cite{dun97};
such a process tends to be more inclusive at low
$m_{c\overline{c}}$ and exclusive at large $m_{c\overline{c}}$,
so we expect that exclusive $B\rightarrow \psi_g K^{(*)}$
will be favoured in the vicinity of 4 GeV.
Quantitative estimates are model dependent and beyond the scope of
this study.

\section{Hybrid decays}

An important feature of hybrid decays in at least flux--tube or bag models
is that decays to two mesons with the same spatial wave function are
suppressed.
This selection rule~\cite{page97sel} is expected to be broken for light
flavours and less so for heavy flavours~\cite{cp95}.
In the case of $\psi_g$, decays to $D^{(*)} D^{(*)}$
are suppressed and the sum of the widths is predicted to be $1-10$ MeV
depending on $J^{PC}$ of the hybrid~\cite{cp95}.
The decays  $1^{-+}\rightarrow \pi\pi,\;
\eta^{(')}\eta^{(')}$ are also suppressed \protect\cite{iddir}.
However, the dissimilar
nature of $\eta_c$ and $\eta^{(')}$ suggests that the decay
$1^{-+}\rightarrow \eta_c\eta^{(')}$ should not be
impeded sizably.

The above selection rule would
be broken if the hybrid states mix with conventional
excitations of $c\overline{c}$.
Hybrid states with exotic $J^{PC}$ are particularly interesting as they
cannot mix with excited $c\overline{c}$ conventional states and, if below 4.3
GeV in mass, will feed $B\rightarrow K+$ ``non-charm".
States with conventional $J^{PC}$  on the other hand can mix with
excited states of the same $J^{PC}$ and thereby
``leak" into  $D^{(*)} D^{(*)}$  final states.
In particular it has  been suggested \cite{ono, cppsi}
that $\psi(4040)$ and $\psi (4160)$ are strong mixtures of
$\psi_{3S}(4100)$ and $\psi_g(4100)$.
In addition, hybrid charmonia may mix with glueballs.  Such a mixing would
enhance the production of light hadrons.

For those $\psi_g$ that mix negligibly with conventional charmonia and have a
mass of $<$ 4.3 GeV, the prominent decays will be either by cascade
$\psi_g [\equiv c \overline c g]  \rightarrow (gg) + (\psi,\eta_c, \ldots )$
or by annihilation  $\psi_g(C=+) \rightarrow  (gg) \rightarrow $ light
hadrons. These are at the same order in $\alpha_s$.  The decay $\psi_g
\rightarrow$ light hadrons is expected to be favoured at least for $C=+$ states
for the following reason.

A measure of the relative importance of the cascade width compared to the
annihilation width may be provided by
$\Gamma (\psi^\prime \rightarrow \psi \pi\pi) \simeq {\cal O}$(0.1 MeV) versus
$\Gamma
(\eta^\prime_c\rightarrow$ light hadrons) $\simeq
\Gamma(\eta_c\rightarrow$ light hadrons) $\times \Gamma^{ee}
(\psi^\prime)/\Gamma^{ee} (\psi) \simeq {\cal O}(5 {\rm\;MeV})$.
The $\psi^\prime \rightarrow \psi \pi\pi$  gives information about
$\psi^\prime \rightarrow \psi gg$, while $\eta^\prime_c\rightarrow$ light
hadrons informs about
$\eta^\prime_c\rightarrow gg$.  Both processes are ${\cal O}(\alpha_s^2)$ in
rate.  Those rates suggest what to expect for cascade and annihilation decays
of charmed hybrids. The rates of
$\psi_g \to (c\overline{c}) +$ light hadrons and
$\psi_g(C=+) \to  $light hadrons are both down by one power in $\alpha_s$.
Ignoring differences in
wave function overlaps  we roughly estimate
$\Gamma(\psi_g \to (c\overline{c}) +$ light hadrons)
$\sim {\cal O}(0.5 {\rm \;MeV})$
and $\Gamma(\psi_g(C=+) \to  $light hadrons) $\sim {\cal O}(20 {\rm \;MeV})$.

The light hadron production rate from $\psi_g$ decays with $C= -$ is expected
to be suppressed by one power of $\alpha_s$ with regards to
$\psi_g(C=+)$ decays.
Note that the production rate of conventional charmonia $(c\overline c)$
from either $\psi_g(C=+)$ or $\psi_g(C=-)$ decays is expected to be of the
same order in $\alpha_s$ and thus similar.
The charge conjugation $(C=\pm )$ of the produced conventional
charmonium is expected to be the same as that of the parent $\psi_g$ in
hadronic
decays, since two gluons $(C= +)$ are emitted in the lowest--order process.
That
may prove
useful in searching for and classifying $\psi_g$ states.

Some decays are forbidden by simple conservation of quantum numbers;
for example, $(J=0) \not\to (J=0) + \gamma$ by angular momentum
conservation.
Similarly, the $D^{(*)} D^{(*)}$ final states are forbidden by
$P$ and/or $C$ conservation for the $J^{PC}=0^{+-}$ exotic hybrid.
Thus, if the mass of the $0^{+-}$ is sufficiently low, it will be seen only in
light hadrons or perhaps also in hidden charm decay modes (see Table 1).

There is an interesting possibility if light hadrons such as
$\eta^{(')}$ or $\omega$ contain $c\overline c$ in their Fock
states~\cite{lightcc}. Charmed hybrids that do not mix could decay into $(\psi
,\eta_c, \chi_c, h_c,\ldots)+(\eta^{(')}, \omega)$ via a
Zweig-allowed 4-charmed intermediate state. That amplitude would of course
interfere with the traditional amplitude governing hidden charmonia production.

\section{Experimental signatures and search strategies}

Some information useful in the search of $\psi_g$ is listed in
Table \ref{t1}.
Not only the exotic $\psi_g$ with $J^{PC}=0^{\pm -}, 1^{-+}, 2^{+-}$ will
involve unique characteristics, even the non--exotic $\psi_g$ that
mix negligibly with conventional charmonia will have striking signatures
(corresponding tables could be produced easily).
There are three major categories of decay: (a) open charm, (b) hidden charm,
and (c) light hadrons. Light hadronic modes that involve one or more $K
\overline K$ pairs (more generally $s\overline s$ pairs) could also be searched
for.
In general, we suggest that a dedicated study of
$B\to \psi_g X_s$, where $X_s$ is light hadron(s) with total strangeness
$= +1$, be made as follows:
\begin{itemize}
\item[(i)]
 $\psi_g\to D^{(*,**)} D^{(*,**)}$. In addition to a search for $\psi_g$, the
 $D^{(*,**)} D^{(*,**)}$ system
 should be studied to seek evidence of $\psi(4040; 4160)$,
 and other excited ($c \overline c$) states (see section 4).
 On general grounds we advocate measuring the
 $J^{PC}$ dependence of charm pair production by these channels.  Note that
such
 channels  feed a ``wrong sign"
$\overline B[b\bar{q}]\rightarrow \overline{D}[\bar{c}q'] +
 \ldots$ charm production that has been observed by several
 experiments~\cite{wrongD},
 and so relevant data may already be at hand.
\item[(ii)]
 $\psi_g\to (c\bar c) +({\rm light\;hadron(s)},\gamma)$, where $(c\bar c)$ is
 a conventional charmonium $\psi,\eta_c,\chi_c,h_c,\ldots$.
 The light hadronic system is likely to have $C=+$ and zero isospin.
 Other quantum numbers, however, should also be
 studied (see Table 1).
\item[(iii)]
 $\psi_g\to$ light hadrons. For examples, consult the last column of Table 1.

\end{itemize}

When $X_s =  K^{(*)}$, it has definite momentum in the $B$
rest frame. Thus
careful studies of $K^{(*)}$ momentum spectra may establish excesses beyond
what is
expected from other sources.

We recommend not only to search for $\psi_g$ production, but also for
other gluonic hadrons in $B$ decay.  Significant yields of light hadrons
in $B$ decays could  feed through light
gluonic hadrons. In that scenario the $b\rightarrow c\overline cs$
transition is followed by non--perturbative $c\overline c$ annihilation, such
as multiple gluon exchange between the spectator quark and the closed charm
line \cite{bjorken}. The resulting intermediate state is rich in soft gluons
and light quarks, which can arrange themselves into glueballs, light
hybrids and glue--rich mesons like $\eta^{'}$. The
resulting final-states have lost their charm content.
Some decay modes of glueballs into light hadrons are summarized in the
last column of Table 1.  Predicted $J^{PC}$ and masses of glueballs can be
found in Table~\ref{t2}.

There is also the possibility for
the $K$ system to resonate as $K_g$. The lightest of these
states is predicted to occur $\sim 2$ GeV in mass
\cite{hy2,cl96} which leaves
$\;\raisebox{-.4ex}{\rlap{$\sim$}} \raisebox{.4ex}{$<$}\;3\;$
GeV available for the mass of the $c\overline{c}$ system. Excitation
of $K_g$ is therefore likely only with low mass  charmonia (such as
$\psi,\eta_c$)  or with
$(\eta^{(')}, \omega,\rho, \ldots)$ if they contain
$c\overline{c}$~\cite{lightcc,brodsky}, or with light  hadrons.
A search for $B \to K_g \eta^{(')}$ could be interesting, in light
of the large $B \to K \eta'$.

We note that vertex detectors can utilize the
long lifetime of $B$ and $D$ hadrons to reduce backgrounds, and the excellent
$p/K/\pi$
separation capabilities at $B$--facilities will further improve the
sensitivities.  Full exploration of multibody decays of $b$-hadrons will
require the ability to detect $\pi^0, \eta^{(')}, \gamma$ as well.

\section{Conclusions}

$B$ decays are a fertile ground for searching and
discovering gluonic hadrons, including hybrid charmonia which
may be copiously produced in the process $b\to  c\bar cs$.
Some of them may significantly decay to light hadrons contributing
to $B$ decays to final states without charm. We have studied the
patterns of production and decay of such hybrids, and proposed
experimental search strategies.

\vspace{.5cm}

\noindent {\bf Acknowledgements}

We are grateful to J. Kuti for discussions.
PRP acknowledges a Lindemann Fellowship from the English Speaking Union.
This work was supported in part by the U.S. Department of Energy
under Contract No. DE-AC02-76CH03000.

\newpage

\begin{table}
\begin{center}
\caption{Some possible experimentally accessible final states of
$J^{PC}$ exotic charmed hybrids and glueballs below $D^{**} D$ threshold.
Note that open charm modes of $\psi_g$ may be suppressed by a selection
rule \protect\cite{page97sel}. For hidden
charm modes, the charmonia tend to have the same $C$ as that of the parent
$\psi_g$. The light hadron modes are expected to be enhanced for $\psi_g$
with $C = +$. See the main text for details.
Decays to $p\bar{p}\{\pi,\eta^{(')},\omega,\rho,\phi\}$ are allowed for
all states listed. }
\label{t1}
\vspace{0.2in}
\begin{tabular}{|c|c|c|c|}
\hline
$J^{PC}$ & Open charm & Hidden charm & Light hadrons \\
\hline
\hline
$0^{+-}$ & Quantum  & $J/\psi \{f_{\{0,1,2\}},(\pi\pi)_S\}$ &
 $a_{\{0,1,2\}} \rho;\; a_{\{1,2\}}\{b_1, \gamma\}$ \\
& numbers & $h_c\eta;\; \eta_c h_1$&
 $b_1 \pi;\; h_1 \eta^{(')}$ \\
& forbid  &
$\chi_{c0}\omega$&$\{(\pi\pi)_S, f_0\}\{\omega,\phi\}$\\
& $D^{(*)} D^{(*)}  $  &
$\chi_{c\{1,2\}}\{\omega,h_1,\gamma\}$&
$f_{\{1,2\}}\{\omega,h_1,\phi,\gamma\}$\\
\hline
$0^{--}$ & $D^* D$&
$h_c (\pi\pi)_S$&
$a_{\{0,1,2\}} b_1;\; a_{\{1,2\}}\{\rho, \gamma\}$ \\
   &   &$J/\psi\{f_{\{1,2\}},\eta^{(')}\} $&
$ \rho\pi$\\
    &           &
$\chi_{c0}h_1;\;\eta_c\{\omega,\phi\} $&
$f_0 h_1;\;\eta^{(')}\{\omega,\phi\}$\\
    &           &
$\chi_{c\{1,2\}}\{\omega,h_1,\gamma\} $&
$f_{\{1,2\}}\{\omega,h_1,\phi,\gamma\} $\\
\hline
$1^{-+}$ & $D^* D$,  $D^* D^*$ &
$\chi_{c\{0,1,2\}} (\pi\pi)_S$&
$a_{\{0,1,2\}} a_{\{0,1,2\}};\;a_{\{1,2\}}\pi$\\
         &          &
$\eta_c\{f_{\{1,2\}},\eta^{(')}\} $ &
$f_{\{0,1,2\}} f_{\{0,1,2\}};\;f_{\{1,2\}}\eta^{(')}$\\
         &           &
$\chi_{c\{1,2\}}\eta$&
$\{\rho,\gamma\}\{\rho,b_1\};\;b_1b_1$\\
    &           &$\{h_c,J/\psi\}\{\omega,h_1,\phi,\gamma\} $&
$\{\omega,h_1,\phi,\gamma\}\{\omega,h_1,\phi,\gamma\}$\\
\hline
$2^{+-}$ & $ D^* D$, $D^* D^*$     &
$\{h_c,J/\psi\} \{f_{\{0,1,2\}},(\pi\pi)_S\}$&
$a_{\{0,1,2\}}\{\rho,b_1,\gamma\} $
\\
& & $\{h_c,J/\psi\}\eta^{(')} $ &
$\{\rho,\gamma,b_1\}\pi $\\
& & $\{\eta_c,\chi_{c\{0,1,2\}}\} \{\omega,h_1,\phi,\gamma\} $ &
$\{\eta^{(')},f_{\{0,1,2\}}\} \{\omega,h_1,\phi,\gamma\}$\\
\hline
\hline
 \end{tabular}
\end{center}
\end{table}

\begin{table}
\begin{center}
\caption{Glueball masses in GeV in the $3 - 4.5$ GeV mass range
accessed by $B\rightarrow K^{(\ast)}+$glueball, according to lattice gauge
theory \protect\cite{ukqcd}. The $0^{--}$ glueball
mass is
poorly determined. No $J^{PC}$ exotic glueballs are expected below 3 GeV.}
\vspace{0.2in}
\label{t2}
\begin{tabular}{|c|c|c|c|c|c|c|c|c|c|}
\hline
$J^{PC}$ &
$1^{+-}       $&$2^{-+}       $&$3^{++}  $&$1^{++}$&$2^{--}$&$1^{--}
   $\\
Mass     &
$2.9\pm 0.3$&$3.0\pm 0.2$&$3.9\pm 0.5  $&$
4.0\pm 0.3$&$4.0\pm 0.4$&$4.6\pm 0.5$\\
\hline
$J^{PC}$ &
$1^{-+}     $&$0^{+-}           $&$ 2^{+-}       $ & & & \\
Mass     &
$\lapprox 4.1$&$\lapprox 3.7    $&$3.9\pm 0.7  $ & & & \\
\hline
\end{tabular}
\end{center}
\end{table}

\end{document}